\documentclass{aa}
\usepackage{graphics,epsfig,astron,amssymb,amsmath,array}
\title{ASCA and other contemporaneous observations of the blazar B2\,1308+326}
\author{D.~Watson\inst{1}
   \and N.~Smith\inst{2}
   \and L.~Hanlon\inst{1}
   \and B.~McBreen\inst{1}
   \and F.~Quilligan\inst{1}
   \and M.~Tashiro\inst{3}
   \and L.~Metcalfe\inst{4}
   \and P.~Doyle\inst{5}
   \and H.~Ter\"{a}sranta\inst{6}
   \and A.~Carrami\~{n}ana\inst{7}
   \and J.~Guichard\inst{7}
}
\thesaurus{03(11.02.2 B2 1308+326; 11.17.4 B2 1308+326; 13.25.2; 13.18.1; 12.07.1)}
\institute{Department of Experimental Physics, University College Dublin, Belfield, Dublin 4, Ireland
    \and Department of Applied Physics and Instrumentation, Cork Institute of Technology, Cork, Ireland
    \and Department of Physics, University of Tokyo, Hongo, Bunkyo-ku, Tokyo, 113-0033, Japan
    \and ISO Data Centre, Astrophysics Division, Space Science Department of ESA, Villafranca del Castillo, P.O. Box 50727, 28080 Madrid, Spain
    \and Physics Department, National University of Ireland, Cork
    \and Mets\"{a}hovi Radio Observatory, FIN-02540 Kylm\"{a}l\"{a}, Finland
    \and Instituto Nacional de Astrof\'{\i}sica Optica y Electr\'onica, Luis Enrique Erro 1, Tonantzintla, Puebla 72840, M\'{e}xico}
\offprints{dwatson@bermuda.ucd.ie}
\date{Received / Accepted }

\begin{document}
\maketitle


\begin{abstract}
The high redshift ($z=0.997$) blazar \object{B2\,1308+326} was observed
contemporaneously at x-ray, optical and radio wavelengths in June 1996.
The x-ray observations were performed with ASCA.
The ASCA results were found to be consistent with reanalysed data
from two earlier ROSAT observations. The combined ASCA and ROSAT
data reveal an x-ray spectrum that is best fit by a broken
power law with absorber model with photon spectral indices of
$\Gamma_{\rm soft}=3.4_{-1.1}^{+5.1}$ and
$\Gamma_{\rm hard}=1.63_{-0.09}^{+0.10}$ and a break energy at
$1.1_{-0.4}^{+0.4}$\,keV in the rest-frame of the blazar. 
The break in the x-ray spectrum is interpreted, from
the shape of the simultaneous broadband spectral energy distribution,
to be the emerging importance of inverse Compton (IC)
emission which dominates the ASCA spectrum.
 The faint optical state reported for these observations (m$_{\rm
V}=18.3\pm0.25$) is incompatible with the high synchrotron flux previously
detected by ROSAT. The IC emission detected by both ROSAT and ASCA was not
significantly affected by the large change in the synchrotron component.

\ion{Mg}{ii} emission was detected with an equivalent width (W$_{\lambda}$)
of $\sim$\,15\AA, significantly different from previously reported values.
The small and variable W$_{\lambda}$ in \object{B2\,1308+326} may be due
to the highly variable continuum and not intrinsically weak lines
in the source.  A lower limit on the Doppler boost factor calculated from
the contemporaneous data is consistent with expectations for highly
polarised quasars and higher than expected for BL~Lacs.  Absorption
at a level of N$_{\mathrm H}=3.0_{-0.6}^{+2.3}\times10^{20}$\,cm$^{-2}$ was
detected which is in excess of the Galactic value of N$_{\mathrm
H}=1.1\times10^{20}$\,cm$^{-2}$, indicating the possible presence of a
foreground absorber.  A gravitational microlensing scenario cannot therefore
be ruled out for this blazar.  No significant variability on timescales of
hours was detected in the optical or x-ray data.

\object{B2\,1308+326} could be a typical radio-selected BL~Lac in terms of
peak synchrotron frequency and optical and radio variability but its high
bolometric luminosity, variable line emission and high Doppler boost factor
make it appear more like a quasar than a BL~Lac.  It is suggested that
\object{B2\,1308+326} be considered as the prototype of this class of
composite source.

\end{abstract}


\section{Introduction}
The classification of the high redshift ($z=0.997$) source
\object{B2\,1308+326} is uncertain \cite{1993ApJ...410...39G}. It has been
designated a BL~Lac \cite{1991ApJ...374..431S} because of its almost
featureless optical spectrum \cite{1979ApJS...41..689W},
its strong optical variability \cite{1985ApJ...288..718M} and the high degree
of polarization of its optical continuum  \cite{1980ARAA..18..321A}.
As one of the 1~Jy sample of BL~Lacs, \object{B2\,1308+326} is classified as
a Radio-Selected BL~Lac (RBL) \cite{1993A&AS...98..393S}.

However, \object{B2\,1308+326} also exhibits quasar-like properties.
VLBI polarisation images show that the polarised flux from the inner part of
the jet is nearly perpendicular to the direction of the jet which is a
characteristic property of quasars not BL~Lacs \cite{1993ApJ...410...39G}. The
degree of core polarisation from VLBI observations of \object{B2\,1308+326}
is somewhat higher than usual for quasars and at the lower end of the range for BL~Lacs
\cite{1993ApJ...410...39G}. Measurements of the apparent speed of knots
emerging from the core are superluminal ($3.6h^{-1}c$, $8h^{-1}c$,
$21h^{-1}c$) and above the typical speeds observed in BL~Lacs
\cite{1993ApJ...410...39G}. \object{B2\,1308+326} also shows an optical-UV
excess above the IR extrapolation \cite{1989ApJ...340..129B},
and its radio power and morphology are FR\,II
\cite[type\,II]{Fanaroff:1974} in nature  \cite{1992AJ....104.1687K}.
The FR\,II classification is not that unusual among BL~Lacs
\cite{1996ApJ...460..174K}, but an FR\,I morphology would strongly
militate against its being a quasar
\cite{2000ApJ...532..816U,1990AJ....100.1057K}.  The large
bolometric \cite{1996ApJ...463..444S} and 5\,GHz luminosities
\cite{1992AJ....104.1687K} are further indications that it could
be regarded as a quasar.  The term ``blazar'' is particularly
useful for sources like this, encompassing as it does flat
spectrum radio quasars (FSRQs) and BL~Lacs.  It must be noted
that the terms highly polarised quasars (HPQs), optically violently
variable (OVV) quasars, FSRQs (and more recently, quasar-hosted blazar)
are used to refer to blazar subclasses.  The significant distinction
here is between quasar-type and BL~Lac-type blazars and the term FSRQ
is used except in cases where another classification occurs in a work
cited in the text.

Two main approaches have proven to be effective in
constraining models of BL~Lacs;  monitoring their variability
(e.g.\ Georganopoulos~\&~Marscher 1998a\nocite{1998ApJ...506L..11G}),
and studying their broadband spectral energy distributions (SEDs)
(e.g.\ Sambruna~et~al.\ 1996\nocite{1996ApJ...463..444S}, Kubo~et~al.
1998\nocite{1998ApJ...504..693K}, Sambruna~et~al.
1999\nocite{1999ApJ...515..140S}). Rapid flux variations and changes
in the shape of the SEDs of blazars are common (e.g. Pian~et~al.\
1998\nocite{1998ApJ...492L..17P}, Macomb~et~al.
1995\nocite{1995ApJ...449L..99M}, Chiaberge \& Ghisellini
1999\nocite{1999MNRAS.306..551C}).  To place adequate constraints
on the synchrotron emission (e.g. the Doppler boost factor of the
putative jet and the angle between such a jet and the line of
sight) in these sources requires simultaneous observations over
a wide range of frequencies.

It has been suggested that some sources with BL~Lac characteristics are
actually gravitationally microlensed quasars
\cite{1986A&A...157..383N,1990Natur.344...45O}.
The best candidates include \object{AO\,0235+164} \cite{1996A&A...310....1R},
\object{PKS\,0537-441} \cite{1999A&AS..135..477R} and \object{MS\,0205.7+3509}
\cite{1999A&A...345..414W}. \object{B2\,1308+326} has
characteristics intermediate between BL~Lacs and quasars.
Gabuzda~et~al.\ \cite*{1993ApJ...410...39G} have therefore suggested that
\object{B2\,1308+326} may be gravitationally microlensed.  Observation of a foreground
galaxy would lend support to the argument that it is indeed gravitationally lensed.
However high resolution imaging with the HST WFPC2
\cite{1999ApJ...512...88U} did not confirm the detection of spatially
extended emission reported by Stickel~et~al.
\cite*{1993A&AS...98..393S}.  The observations were consistent with
a point source with an upper limit of
m$_{\rm I}\geq26$\,mag\,arcsec$^{-2}$ for any surrounding galaxy.

In the superluminal micro-lensing scenario, the time scale
for rapid variability can constrain source parameters (e.g.\
Lorentz factor and source diameter, McBreen \& Metcalfe
1987\nocite{1987Natur.330..348M}, Gopal-Krishna \& Subramanian
1991\nocite{1991Natur.349..766G}). Such microlensing variations
should be achromatic and therefore detectable with simultaneous
multi-wavelength observations.

Simultaneous x-ray, optical and radio observations were carried
out in order to constrain the nature of the emission in
\object{B2\,1308+326}. These observations and the data reduction
procedures are described in Sect.~2.  Results are presented in
Sects.~3 and 4 and are discussed in Sect.~5 with conclusions in
Sect.~6.
$H_{0}=75$\,km\,s$^{-1}$\,Mpc$^{-1}$ and $q_{0}=0.5$
are assumed throughout.


\section{Observations and data reduction}
\subsection{X-ray observations}
Observations of \object{B2\,1308+326} were made with the ASCA satellite
\cite{1994PASJ...46L..37T} between 15:26 and 23:41\,UT on 10 June 1996,
and between 23:37\,UT on 11 June 1996 and 09:10\,UT on 12 June 1996. There
are four instruments on board ASCA, two Gas Imaging Spectrometers (GIS-S2
and GIS-S3) and two Solid-state Imaging Spectrometers (SIS-S0 and SIS-S1).
The instruments have a well-calibrated energy range of 0.7--10.0\,keV for
the GIS and 0.5--10.0\,keV for the SIS and only data in these energy ranges
were used for spectral fitting. The total exposure time per instrument was
approximately 40\,ks.  The SIS data were recorded in FAINT data mode and
converted to BRIGHT-2 mode \cite{1993ExA.....4....1I} having corrected for
the Residual Dark Distribution (RDD) effect with the FTOOLS script
`Correctrdd'.

Data from the SIS and GIS instruments were screened in a standard way, and
were then reduced using the FTOOLS applications. An extraction region
was defined around the source, with an aperture of 4\arcmin\, for the SIS
detectors and 6\arcmin\, for the GIS. This spatial region was used to
extract the source plus background counts.  An estimate of the background
was derived by taking an annulus around the extraction aperture and
extracting counts from this region. An alternative background estimate was
derived from separate dark-sky observations made by ASCA and results
obtained using both types of background estimates were found to be
consistent. The annulus background was subtracted from the extracted source
plus background counts to obtain a source spectrum.  The resulting spectra
were rebinned to have a minimum of 20 counts per bin in order to ensure the
validity of the Gaussian approximation.

Two observations of the source were made with the ROSAT PSPC \cite{1986SPIE..597..208P}
on 23 June 1991 \cite{1995MNRAS.277..297C,1996A&A...311..384L} and 3 June 1992.
These data were reduced in a standard way using the FTOOLS software.
The response matrix from ROSAT AO-1 was used to analyse the data from the 23
June 1991 observation, and that from AO-2 for the 3 June 1992 observation.
Only the well-calibrated data between 0.1--2.4\,keV were used.


\subsection{Optical and radio observations}

Optical Johnson V-band observations were carried out under photometric
conditions with the 1.23\,m telescope at Calar Alto Observatory, Spain
on the nights of 10 and 11 June 1996 with typical exposure times of 900\,s.
Standard CCD aperture photometry techniques were used to reduce the data.
Absolute calibration of the source was derived using V-band differential
photometry on a calibrated reference star within the CCD field of view
(star~C from Fiorucci~et~al.\ 1998\nocite{1998PASP..110..105F}). Optical
V-band observations made two weeks later with the 1\,m JKT telescope at
La~Palma Observatory showed the source to be of the same magnitude, within
the error of 0.25\,mag.  The relative uncertainty (0.08\,mag.)\ in the optical
magnitude was brought below the absolute calibration uncertainty
(0.25\,mag.), by using differential photometry techniques
\cite{1998A&AS..129..445R}.

Observations were carried out under non-photometric conditions at the
2.12\,m telescope of the Observatorio ``Guillermo Haro'' at Cananea, Son.,
M\'exico on June 16, 1996 using the Landessternwarte Faint Object
Spectrograph and Camera (LFOSC) in
$\sim$5.5\,\AA /pixel mode.  The standard IRAF utility package distributed
by the National Optical Astronomy Observatories was used to reduce the data.
The spectral images were processed to obtain one-dimensional spectra of the
object and the sky by co-adding the counts on each channel along the slit,
through user defined apertures.

The wavelength calibration was performed using the identified lines of the
comparison arcs in order to obtain a dispersion solution. With the derived
coefficients the wavelength calibration of the spectra was then carried out.
Due to the non-photometric conditions during these spectral observations,
the spectrophotometric calibration proved to be unreliable.

The results of radio observations at 22\,GHz and 37\,GHz performed at the
Mets\"{a}hovi Radio Observatory, Finland for the period 1980--2000 are
included here. The observing and data reduction procedures are described in
Ter\"{a}sranta et~al.\ \cite*{1998A&AS..132..305T}.



\section{X-ray Spectral Fits}

The x-ray data from both ASCA observations were compared for
variability above a significance level of $2\,\sigma$.  No variability was
detected and the data were therefore coadded per detector for further analysis.

Data from the four ASCA instruments
were fit individually using the Levenberg-Marquardt algorithm in XSPEC.
Model~A, a power-law with an absorber as defined by Morrison \& McCammon
\cite*{1983ApJ...270..119M} was fit to these data and yielded
acceptable fits and consistent results (Table~\ref{instrumentresults}).

Since the ASCA instruments showed no significant spectral differences, the
binned spectra from all four detectors were fit simultaneously.  The data
were fit to model~A (Table~\ref{instrumentresults}) and to model~B, a broken
power-law with absorber (Table~\ref{obscompare}).  The $f$-test
\cite{1992drea.book.....B} was applied to models~A and B to establish
whether the fit was significantly improved by the addition of the two extra
terms in model~B (Table~\ref{models}).  Model~B did not yield a
significant improvement over model~A in the fit to the ASCA data alone
(Table~\ref{models}).

The ROSAT data from 23 June 1991 were published by Comastri~et~al.
\cite*{1995MNRAS.277..297C}.  The results of a single power-law
with absorber fit (model~A) to this ROSAT data were found to be
compatible with and had a similar reduced $\chi ^{2}$ statistic,
$\chi^2_\nu$ (Table~\ref{instrumentresults}) to that obtained by
Comastri~et~al.\ \cite*{1995MNRAS.277..297C} for the same data and
model.  Although model~A provides an acceptable fit
($\chi^2_\nu=1.13$) to the ROSAT data of 23 June 1991 (R1), the
$f$-test indicates that a broken power-law is a significantly
better fit (Table~\ref{models}).  The spectral break is clearly
detected in this data alone (Fig.~\ref{xspec}).  Data from the
ROSAT observation of 3 June 1992 (R2) is acceptably fit by
model~A;  there is no significant improvement in the fit to this
data using model~B (Table~\ref{models}).

\begin{table}
   \caption[Comparison of fits to data from all x-ray observations of B2\,1308+326]{
        Comparison of fits (using model~A, a simple power-law plus
        absorber) to data from: each ASCA detector individually (S0, S1,
        S2, S3), from the combined ASCA detectors (A), from the ROSAT
        observations of 23 June 1991 (R1) and 3 June 1992 (R2) and from
        both ASCA and ROSAT, fit simultaneously (A+R) .  $\rm N_{\rm H}$
        is the equivalent hydrogen column, $\Gamma$ is the photon index,
        F$_{1\mathrm{keV}}$ is the flux in photons/keV/cm$^{2}$/s at
        1\,keV, $\chi^2$/DOF is the value of $\chi^2$ for that fit
        with DOF degrees of freedom and N.~H.~Prob. is the probability
        of the $\chi^2$ obtained being at least as large as it is if the
        data come from a parent distribution described by the model.
        Ranges in parentheses are 90\% confidence intervals for 1
        parameter of interest.}
   \label{instrumentresults}
   \begin{tabular}{@{}l c c c c@{}}
      \hline\hline
          & $\rm N_{\rm H}$ & $\Gamma$ & F$_{1\rm keV}$ & $\chi^2$/DOF\\
          & ($10^{20}$cm$^{-2}$) & & ($\times10^{-4}$) & N. H. Prob.\\
      \hline
      S0 & 0.0 & 1.45 & 1.94 & 46/45 \\
        & (0.0--11.6) & (1.28--1.71) & (1.67--2.55) & 0.45 \vspace{6pt}\\
      S1 & 0.4 & 1.38 & 1.82 & 29/38 \\
        & (0.0--14.6) & (1.18--1.74) & (1.52--2.60) & 0.84 \vspace{6pt}\\
      S2 & 24.5 & 2.21 & 2.94 & 37/49 \\
        & (0.0--102.4) & (1.59--3.24) & (0--9.20) & 0.90\vspace{6pt}\\
      S3 & 0.0 & 1.83 & 2.42 & 67/61 \\
        & (0.0--22.3) & (1.52--2.23) & (1.82--3.59) & 0.27\vspace{6pt}\\
      A & 0.0 & 1.75 & 1.99 & 205/202 \\
        & (0.0--2.5) & (1.66--1.84) & (1.87--2.16) & 0.43\vspace{6pt}\\
      R1 & 1.35 & 2.01 & 2.19 & 34/30 \\
    & (0.84--1.94) & (1.76--2.27) & (1.99--2.39) & 0.29\vspace{6pt}\\
      R2 & 0.81 & 1.75 & 1.89 & 10/12 \\
    & (0.00--2.15) & (1.29--2.30) & (1.61--2.17) & 0.66\vspace{6pt}\\
      A+R & 1.36 & 1.88 & 2.15 & 250/247 \\
        & (1.14--1.58) & (1.80--1.96) & (2.05--2.25) & 0.43\vspace{6pt}\\
      \hline
   \end{tabular}
\end{table}

The energy of the spectral break is at the low energy end of the SIS
detector range and outside that of the GIS.  The ROSAT observation of 3
June 1992 unfortunately does not constrain the models very well.  It is not
surprising therefore that there is no significant improvement in the fit for
either of these observations using a broken power-law model. Data from both
ROSAT observations were fit with model~B and the results were found to be
consistent with the results from the fit of model~B to the ASCA data
(Table~\ref{obscompare}).  No variation was found at the level of
$95\%$ significance, in the 0.5--2.4\,keV flux between any of the
observations.

The consistent results indicate that the x-ray spectrum had not altered
substantially between the ROSAT and ASCA observations (see the next section
for further discussion on this point). The combined ROSAT and ASCA data were
therefore fit simultaneously, allowing the normalization parameter to vary
independently between the two ROSAT observations and the combined ASCA data,
but constraining the other parameters to be the single best-fit for all
datasets.  The flux normalizations were left free to vary independently to
allow for any residual systematic differences between the instruments.

Model~B was the best fitting model to the combined ASCA and
ROSAT data and was used for all further analysis. Although a
discrepancy of $\Delta\Gamma\sim0.4$ has been noted in the spectral indices
reported for observations with ASCA SIS and ROSAT PSPC
\cite{1999MNRAS.307..611I}, the detection of the spectral break does not
depend on differences between the ROSAT and ASCA spectral indices since
the break is clearly detected in the R1 data alone. The results of a comparison
between models~A and B for all the x-ray observations are summarised in
Table~\ref{models}.


\section{Results}
\begin{figure*}
  \begin{center}
  \epsfig{file=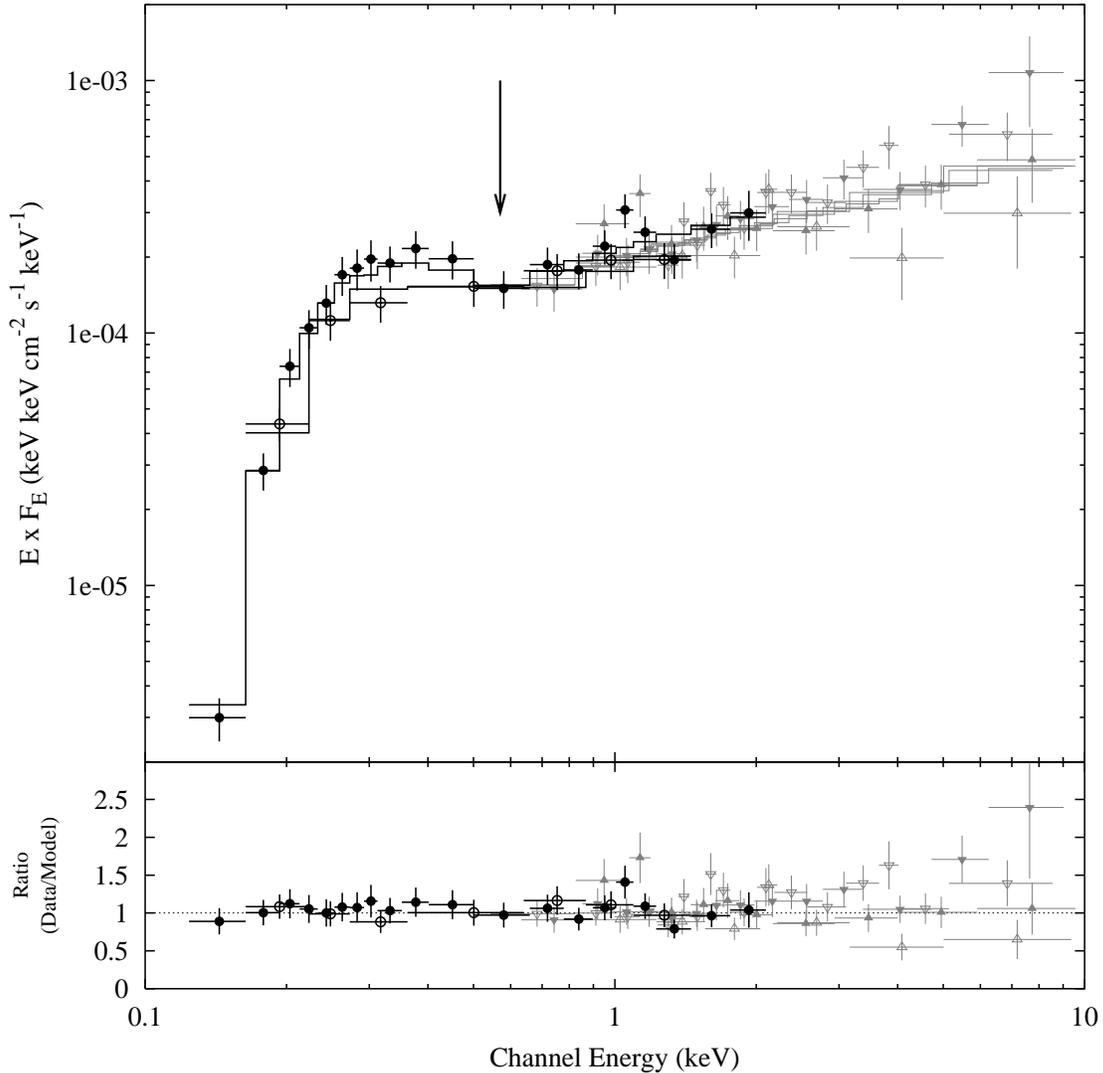,width=1.67\columnwidth,clip=}
  \end{center}
  \caption[Observer frame E$\times$F$_{\mathrm{E}}$ x-ray spectrum of B2\,1308+326]{
       Best fit (observer frame) E$\times$F$_{\mathrm{E}}$ spectrum
           to joint ASCA and ROSAT data with data/model ratio.  The break at
           0.57\,keV is indicated.  Data from the ASCA instruments are
           plotted in grey (S0~`$\triangledown$', S1~`$\blacktriangledown$',
           S2~`$\vartriangle$' and S3~`$\blacktriangle$').  ROSAT
           observations are plotted in black (R1~(23 June 1991)~`$\bullet$',
           R2~(3 June 1992)~`$\circ$'.)}
  \label{xspec}
\end{figure*}

The deconvolved data from all instruments with the best-fit model and the
data/model ratio are plotted in Fig.~\ref{xspec}.

\begin{table}
\begin{minipage}{\columnwidth}
   \caption[Results from broken power-law fit to x-ray observations of B2\,1308+326]{
        Results from best-fit broken power-law plus absorber model (B),
            to data from different observations; R1 and R2 are the ROSAT
            observations of 23 June 1991 and 3 June 1992 respectively; A
            represents the simultaneous fit to the data from the ASCA
            observations of 10--12 June 1996; A+R is a simultaneous fit to
            all the ROSAT and ASCA data.  Ranges in parentheses are 90\%
            confidence intervals for 1 parameter of interest.  $\Gamma1$
            and $\Gamma2$ are the photon spectral indices below and above
            the break energy respectively.}
   \label{obscompare}
   \begin{tabular}{@{}l c c c c@{}}
      \hline\hline
          & $\rm N_{\rm H}$ & $\Gamma$1 & Break & $\Gamma$2 \\
          & ($10^{20}$\,cm$^{-2}$) & (soft) & (keV) & (hard) \\
      \hline
      R1 & 8.4 & $\geq10.0$ & 0.38 & 2.1 \\
              & (6.35--8.69) & (3.04--) & (0.35--0.43) & (1.7--2.4) \\
      \footnote{$\Gamma 1$ is unconstrained for the ASCA and R2 data, since model~B is not a significantly better fit than model~A for these data alone (see Table~\ref{models}).}R2 & 7.1 & 10.0 & 0.33 & 2.32 \\
              & (5.1--7.6) & (--) & (0.28--0.42) & (1.58--2.49) \\
      $^a$A & 2.78 & 10.0 & 0.45 & 1.62 \\
           & (0.0--9.7) & (--) & (0.21--0.56) & (1.54-1.77) \\
      A+R & 3.0 & 3.4 & 0.57 & 1.63 \\
          & (2.4--5.3) & (2.3--8.5) & (0.36--0.77) & (1.54--1.73) \\
      \hline
   \end{tabular}
\end{minipage}
\end{table}

\begin{table}
   \caption[Comparison of fits to x-ray observations of B2\,1308_+326]{
        Fit comparisons for models~A, power-law plus absorber and B,
            broken power-law plus absorber. N.~H.~Prob. is the probability
            of the $\chi^2$ obtained being at least as large as it is if the
            data come from a parent distribution described by the model.
            The $f$-test Prob. refers to the probability derived from the
            $f$-test of the hypothesis that an additional component (i.e.
            spectral break) yields an improved description of the data.  R1
            and R2 are the ROSAT observations of 23 June 1991 and 3 June
            1992 respectively; A represents the ASCA observations of 10--12
            June 1996; A+R is a simultaneous fit to all the ROSAT and ASCA
            data.  The spectral break is not significant in the ASCA or R2
            data individually, but is significant in R1 alone and in the A+R
            simultaneous fit.}
   \label{models}
   \setlength{\tabcolsep}{5pt}
   \begin{tabular}{@{}l |l@{}r|l r|r@{}}
       \hline\hline
       Obs.& \multicolumn{2}{|c|}{A: abs$\times$pow} & \multicolumn{2}{c|}{B: abs$\times$bknpow} & $f$-test\\
       & N.H. Prob.& $\frac{\chi^{2}}{\mathrm{DOF}}$ & N.H. Prob.& $\frac{\chi^{2}}{\mathrm{DOF}}$ & Prob.\\ \cline{3-3} \cline{5-5}
       \hline
       R1 & 0.286 & 34/30 & 0.707 & 24/28 & 0.994\\
       R2 & 0.657 & 10/12 & 0.657 & 7.7/10 & 0.651\\
       A  & 0.431 & 205/202 & 0.433 & 203/200 & 0.645\\
       A+R & 0.430 & 250/247 & 0.621 & 238/245 & 0.998\\
      \hline
   \end{tabular}
\end{table}

\begin{figure}
\epsfig{file=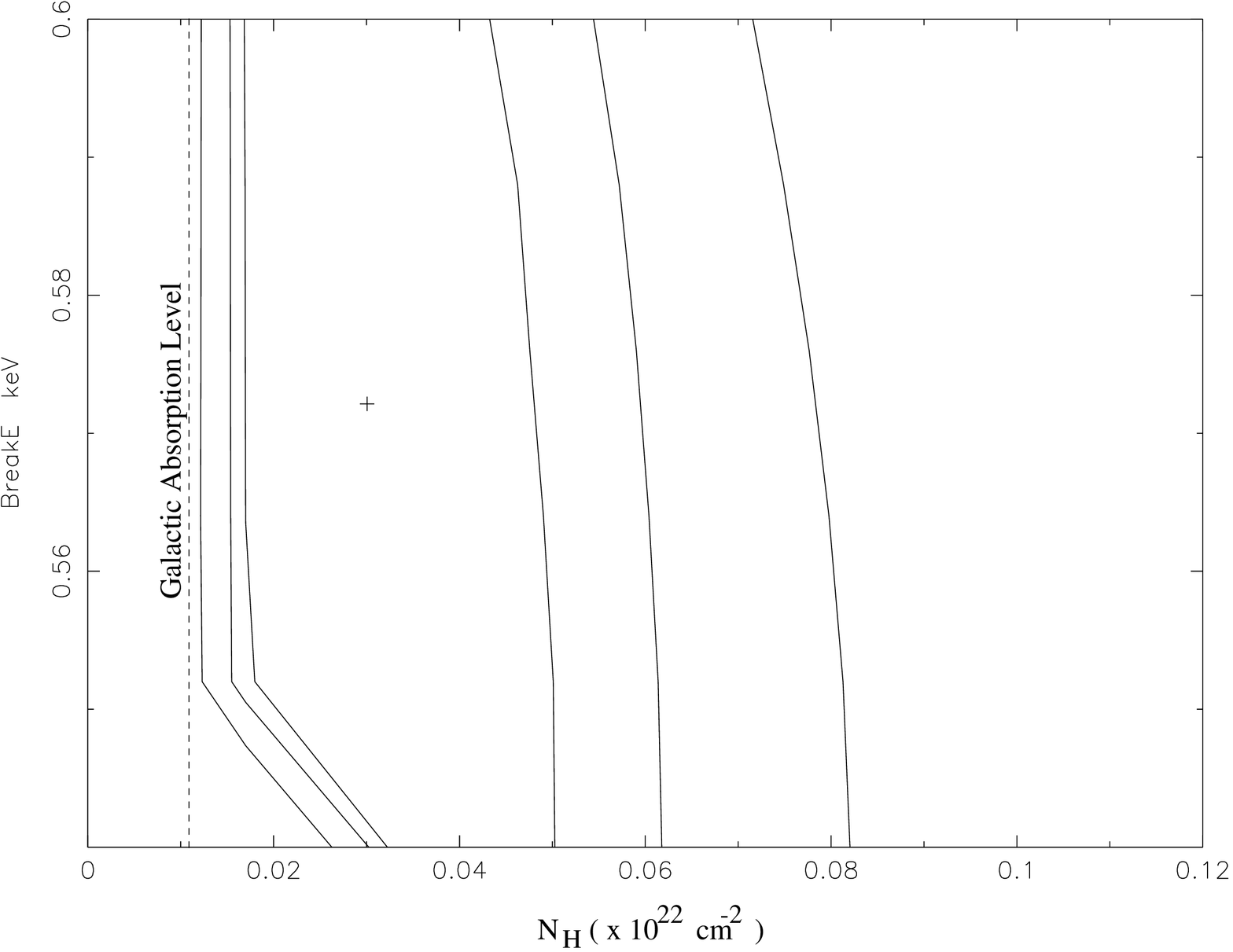,height=\columnwidth,angle=-90,clip=}
\caption[Confidence contours N$_{\rm H}$ and break energy for x-ray data from B2\,1308+326]{
     68\%, 90\% and 99\% confidence contours for two parameters of
         interest; equivalent hydrogenic absorbing column (N$_{\rm H}$) and
         spectral break energy, both in the observer frame.  The dashed line
         represents the equivalent Hydrogen column density level of the
         Galaxy in the direction of \object{B2\,1308+326}
         \protect\cite{1990ARA&A..28..215D}.}
\label{1308contours}
\end{figure}

Confidence contours for the best-fit model for the two parameters of
interest, equivalent Hydrogen absorbing column ($\rm N_{\rm H}$) and the
spectral break energy, are shown in Fig.~\ref{1308contours}.  The soft x-ray
absorption at $3.0\times10^{20}$\,cm$^{-2}$ is above the Galactic level of
$1.1\times10^{20}$\,cm$^{-2}$ given by Dickey \& Lockman
\cite*{1990ARA&A..28..215D}.

\begin{figure}
\begin{minipage}{\columnwidth}
\epsfig{file=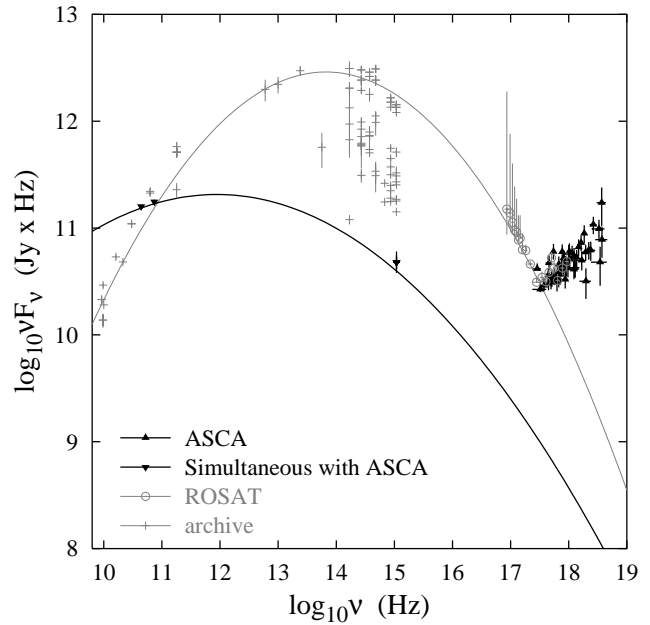,width=\columnwidth,clip=}
\caption[Broadband spectral energy distribution of B2\,1308+326]{
     Rest-frame, $\nu$F$_\nu$ spectral energy distribution of
         \object{B2\,1308+326}. The x-ray data has been corrected for
     absorption using the best-fit parameters (Table~\ref{obscompare}).
     The break at $2.8\times10^{17}$\,Hz is clearly seen in the middle
     of the x-ray data.  Data contemporaneous with the ASCA observation
     are plotted in black.  Non-simultaneous data are plotted in grey.
     Archive data are from NED\protect\footnotemark[1].  Two parabolic
     fits to the synchrotron component are plotted, the black one is the
     best fit parabola to the contemporaneous data below the break, the
     grey parabola is a fit to the highest archival flux values in each
     band (from NED) and the ROSAT data below the break.  The shift in
         frequency of the synchrotron peak is apparent.}
\label{broad}
\end{minipage}

\end{figure}

The rest-frame SED is plotted in Fig.~\ref{broad}.  The break in
the ROSAT spectrum is clearly evident at 1.1\,keV
($2.8\times10^{17}$\,Hz).  It is generally accepted that the
two broad emission components that characterise blazar SEDs are
due to synchrotron emission at lower energies and inverse Compton
(IC) emission at higher energies
\cite{astro-ph/0001410,astro-ph/0005066,1996ApJ...463..444S,1996ApJ...459..169P}
The SED of \object{B2\,1308+326} shows the synchrotron component
peaking between $10^{12}$\,Hz and $10^{14}$\,Hz with the inverse
Compton emission dominating by $\sim2.8\times10^{17}$\,Hz.
The parabolic fit to the contemporaneous
data indicates that the peak synchrotron frequency is
$\sim10^{12}$\,Hz (Fig.~\ref{broad}).
 \label{SEDchange}Although no
variation in the 0.5--2.4\,keV spectra was detected between the
observations made with ROSAT and ASCA, examining the SED
(Fig.~\ref{broad}) leads to the conclusion that the ROSAT
synchrotron flux is incompatible with the radio and optical data
obtained contemporaneously with ASCA.  The effect of the
absence of the synchrotron flux on the R1 spectrum was quantified by
extrapolating the best-fit ROSAT synchrotron power-law past the spectral
break to 2.4\,keV. The 0.5--2.4\,keV flux of this synchrotron power-law
alone was subtracted from the composite synchrotron-IC R1 0.5--2.4\,keV
flux.  The resulting (IC) flux was still fully consistent with the ASCA
0.5--2.4\,keV flux.  This indicates that the synchrotron-IC break
occurred at lower energies during the ASCA observation than
during R1 and that the ASCA spectrum is almost entirely dominated
by an IC component that has not varied substantially at these
energies despite the large change in the synchrotron flux.
Chiaberge \& Ghisellini \cite*{1999MNRAS.306..551C} have modelled
synchrotron--self-Compton emission in blazars and successfully
reproduced the SED of \object{Mkn\,421} in different flux states
\cite{1995ApJ...449L..99M}.  Their model allows a peak synchrotron
variation of nearly two orders of magnitude that leaves the
low-frequency end of the IC component largely unaffected,
consistent with the SED observed here and in \object{Mkn\,421}
\cite{1995ApJ...449L..99M}.  Major variability of the IC component
would however be expected at $\gamma$-ray energies because of the large
change in the synchrotron emission.  Observations at $\gamma$-ray energies
can strongly constrain synchrotron-IC models \cite{1999ApJ...526L..81M}.  A
parabolic fit to the highest archival fluxes in each waveband and the ROSAT
data below the spectral break yields a peak synchrotron frequency of
$\sim10^{14}$\,Hz. This indicates that the peak synchrotron
frequency of \object{B2\,1308+326} may decrease with flux, an
important indicator in deciding the nature of the inverse Compton
emission in this source \cite{1999ApJ...515L..21B}.  This type of
spectral break and the possibility of increasing synchrotron peak
frequency with increasing flux have also been observed with
{\em BeppoSAX} in the blazar \object{S5\,0716+714} \cite{1999A&A...351...59G}.

\footnotetext[1]{The NASA/IPAC Extragalactic Database (NED) is
operated by the Jet Propulsion Laboratory, California Institute of
Technology, under contract with the National Aeronautics and Space
Administration.}

The ASCA lightcurve yields no evidence for variability above
$7.5\times10^{-14}$\,erg\,cm$^{-2}$\,s$^{-1}$ on timescales of hours. 
V-band optical photometry from both Calar Alto observations yields a
magnitude of $18.3\pm0.25$.  A limit on the variability of the source of
0.08\,mag.\ was found using differential photometry.  Therefore, no
constraints can be placed on the superluminal gravitational lensing 
scenario in this AGN because of the absence of variability.

\begin{figure}
\epsfig{file=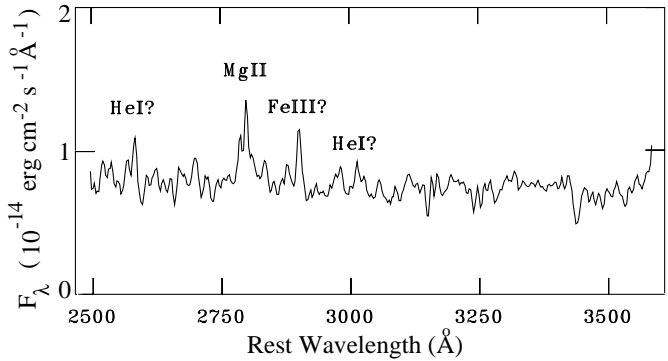,width=\columnwidth,clip=}
\caption[Optical spectrum of B2\,1308+326]{
     Redshift ($z=0.997$) corrected optical spectrum.  The absolute flux
     calibration is unreliable due to non-photometric conditions during
     the observation, but is included for completeness.  The
     \ion{Mg}{ii} emission feature is identified.}
\label{optical}
\end{figure}

The optical spectrum (Fig.~\ref{optical}) observed on 16 June 1996
is intermediate between the extreme spectra previously observed
\cite{Miller:78,1993A&AS...98..393S}.
The equivalent width of the \ion{Mg}{ii} line was $\sim$15\,\AA.  The absolute
flux calibration for this spectrum is unreliable due to non-photometric
conditions during the observation;  it was therefore not possible to obtain a
continuum level or line fluxes from the ASCA observations.

\begin{figure}
 \epsfig{file=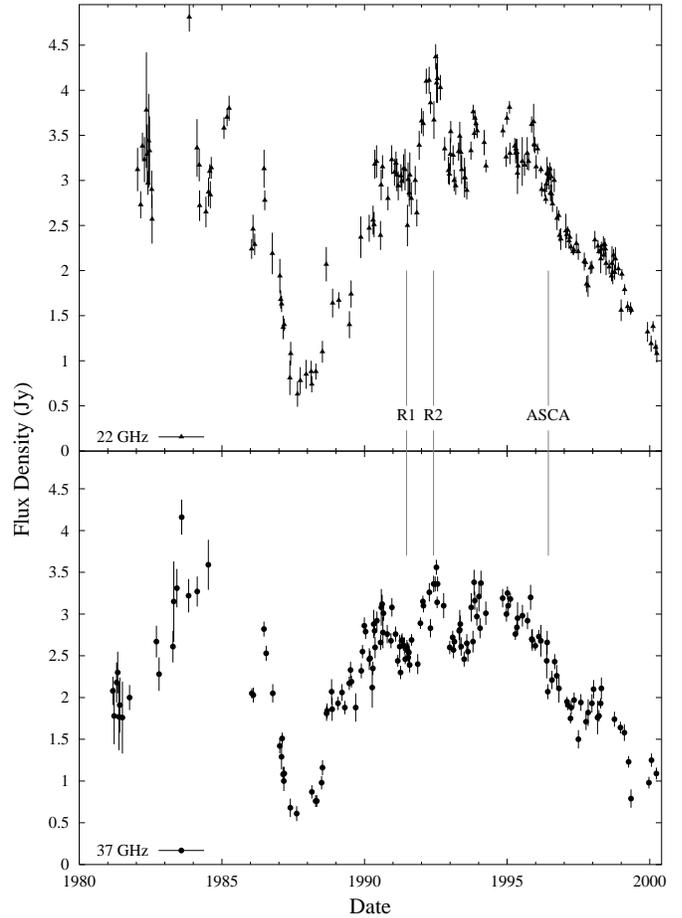,width=\columnwidth,clip=}
 \caption[Radio lightcurve of B2\,1308+326 at 22\,GHz and 37\,GHz]{
      1980--2000 radio lightcurve at 22\,GHz (top) and 37\,GHz (bottom)
      obtained at the Mets\"{a}hovi Radio Observatory.
      The times of the ASCA and ROSAT (R1 and R2) observations are indicated.}
 \label{radio}
\end{figure}

The results of the radio observations (Fig.~\ref{radio}) over 20 years
indicate that \object{B2\,1308+326} was
in an intermediate radio flux state during these observations.

\section{Discussion}

Although BL~Lacs were originally classified as XBLs or RBLs
depending on the energy band in which they were first discovered,
recent work has focused on determining the physical distinctions
between the two types
\cite{1995ApJ...444..567P,1997MNRAS.289..136F,1998ApJ...506..621G}
 and their relation to the other subclass of blazar, FSRQs
\cite{2000ApJ...533..650S,1997ApJ...487..536S,1998MNRAS.299..433F,1998MNRAS.301..451G}.
The spectral energy distributions of XBLs and RBLs differ markedly, with the
peak in synchrotron output generally occurring in the x-ray range for the
former and the infrared range for the latter \cite{1995ApJ...444..567P}.
This property has allowed BL~Lacs to be classified as low-frequency peaked
BL~Lacs (LBLs, roughly the same sample as RBLs) and high-frequency peaked
BL~Lacs (HBLs, approximate to XBLs).  Such differences may arise due to changes in the
dominant radiation mechanism occurring as a function of both the angle
between the line of sight and the bulk velocity \cite{1989ApJ...340..181G},
although it has been demonstrated, assuming certain conditions, that
the complete range of properties observed in blazars are difficult to
account for solely by a change in orientation of the jet
\cite{1996ApJ...463..444S}.  On the other hand, XBLs and RBLs have different
variability timescales, a fact that is very naturally explained with the
orientation hypothesis \cite{1998A&A...329..853H}.
The recent RGB sample revealed BL~Lacs intermediate between XBLs and RBLs,
indicating that there is no bi-modal distribution in BL~Lacs and that they
span peak synchrotron frequencies continuously from IR to x-ray
wavelengths \cite{1999ApJ...525..127L}.

One hypothesis connecting FSRQs and BL~Lacs is that they are high and low
luminosity radio galaxies respectively that are highly-relativistically
beamed towards the observer \cite{1992MNRAS.257..404P}.  A more recent
theory connects FSRQs, LBLs and HBLs in a sequence of increasing peak
synchrotron frequencies and decreasing bolometric luminosity
\cite{1998MNRAS.299..433F,2000ApJ...533..650S};  however,
roughly 25\% of FSRQs in the Deep X-ray Radio Blazar Survey
\cite{1998AJ....115.1253P} have HBL-like radio--x-ray broadband spectral
indices  ($\alpha_{\rm RX}$), indicating that the synchrotron peak frequencies
($\nu_{\rm peak}$) for FSRQs reach energies at least as high as those of HBLs,
since $\alpha_{\rm RX}$ is strongly correlated with $\nu_{\rm peak}$
\cite[and references therein]{1998MNRAS.299..433F}.  In this theory, blazars
with strong thermal emission or strong emission lines (FSRQs) should have
soft x-ray spectra dominated by inverse Compton (IC) emission produced by
jet electrons scattering photons produced externally to the jet (external
Compton, EC model).  Blazars with intrinsically weak emission lines
(BL~Lacs) are expected to have their IC component dominated by jet electrons
scattering photons produced in the jet via synchrotron emission (synchrotron
self-Compton, SSC model). B\"{o}ttcher
\cite*{1999ApJ...515L..21B} has postulated that EC-dominated spectra
have peak synchrotron frequencies that decrease with increases in
flux.
 The peak synchrotron frequency of \object{B2\,1308+326} appears to
decrease with decreasing flux.  While this synchrotron peak shift behaviour
is contrary to the trend found in samples of blazars
\cite{1998ApJ...504..693K,1998MNRAS.299..433F} it is not at all unusual in
flares detected in individual sources (e.g.\ Catanese~et~al.\
1997)\nocite{1997ApJ...487L.143C}.  Without IR photometry it is not possible to
determine precisely the peak synchrotron frequency at the time of these
observations.  The optical observations were obtained
simultaneous with, and two weeks after, the ASCA observations and yielded
compatible results on both occasions.  The radio data nearest the ASCA
observations were taken on 4 June 1996 at 37\,GHz, and on 5 and 20 June 1996 at
22\,GHz.  The radio variability during the surrounding months was less than
30\%.  The spectrum would have to deviate significantly from a parabolic
shape in order to peak above $10^{14}$\,Hz and still manage to drop as low
as 18.3\,mag.\ in the optical.  Such a spectral shape cannot be ruled out
without accurate modelling of the synchrotron component, however it is very
likely that the synchrotron peak did occur below
$10^{14}$\,Hz in the faint state observed here, indicating that the
synchrotron peak frequency decreases with decreasing flux and that
SSC is the more likely emission mechanism in this blazar
\cite{1999ApJ...515L..21B}. Simultaneous GeV observations are
required to determine if there is excess IC emission resulting from the EC
component.  Such observations would be useful for testing the model proposed
for \object{PKS\,0548+134} \cite{1999ApJ...527..132M} and the trend detected
by Kubo~et~al.\ \cite*{1998ApJ...504..693K} in a sample of 18 blazars.

The dominance of the IC component in \object{B2\,1308+326} above
$2.8\times10^{17}$\,Hz is evident from Fig.~\ref{broad}. The relatively low
energy of this break is consistent with expectations of LBLs
\cite{1995A&AS..109..267G}, but somewhat exceeds expectations for
FSRQs in the blazar paradigm described by Sambruna~et~al.
\cite*{2000ApJ...533..650S}.  The contemporaneous nature of the observations
rules out distortion of the spectrum due to source variability.

The primary reason for classifying \object{B2\,1308+326} as a BL~Lac rather
than a FSRQ is its largely featureless optical continuum which is one of the
defining features of the BL~Lac class, with rest-frame W$_{\lambda}<5$\,\AA\ for
any line \cite{1996ApJ...463..444S}.  This criterion has been
justifiably criticised as arbitrary
\cite{1992AJ....104.1687K,1996MNRAS.281..425M,1997A&A...325..109S} partly on
the basis that some sources (such as \object{B2\,1308+326}) have
W$_{\lambda}$ that vary substantially across this boundary. Nevertheless,
weak or absent emission lines is a distinguishing property of these sources
\cite{1992MNRAS.257..404P}.  The rest-frame W$_{\lambda}$ of the
\ion{Mg}{ii} line in \object{B2\,1308+326} detected by Miller~et~al.
\cite*{Miller:78} was $<5$\,\AA\ and no lines were observed by Wills \&
Wills \cite*{1979ApJS...41..689W}.  However, an \ion{Mg}{ii}
W$_{\lambda}=18.6$\,\AA\ in the rest-frame was observed in March 1989 by
Stickel~et~al.\ \cite*{1993A&AS...98..393S} in a very low luminosity state
m$_{\rm R}=19.0$\,mag (AB$_{5500{\mathrm\AA}}\sim18.5$\,mag from the
spectrum), comparable in brightness to the observations presented here. The
W$_{\lambda}$ ($\sim15$\,\AA) reported in this paper and by Stickel~et~al.
\cite*{1993A&AS...98..393S} place \object{B2\,1308+326} outside the
traditional limits for BL~Lacs, while previous observations place it
well inside.

The level of the continuum may be the most significant factor in
changing the value of W$_{\lambda}$ for the \ion{Mg}{ii} line and not
variation of the intrinsic line luminosity, since (a) both large
W$_{\lambda}$ measurements were made near low luminosity states
and (b) the line flux of $3.8\times
10^{-15}$\,erg\,cm$^{-2}$\,s$^{-1}$ reported by Miller~et~al.\
\cite*{Miller:78} is within a factor of two of the value of
$5.8\times10^{-15}$\,erg\,cm$^{-2}$\,s$^{-1}$ of Stickel~et~al.\
\cite*{1993A&AS...98..393S} even though W$_\lambda$ differs
substantially between the observations.  No error
was given for these measurements.  Padovani
\cite*{1992MNRAS.257..404P} has shown that in general the low
values of W$_\lambda$ observed in BL~Lacs are due to intrinsically less
luminous line fluxes than in quasars. Sambruna~et~al.
\cite*{2000ApJ...533..650S} find \ion{Mg}{ii} luminosities of
L$_{\ion{Mg}{ii}}\sim$(4--40)$\times10^{43}$\,erg\,s$^{-1}$ for a sample of
three FSRQs, comparable to but higher than \object{B2\,1308+326}, with
L$_{\ion{Mg}{ii}}=1.5\times10^{43}$\,erg\,s$^{-1}$. It therefore appears
that the small values of W$_\lambda$ observed in
\object{B2\,1308+326} are due mostly to the extremely bright
continuum. Unfortunately, it was not possible to obtain a line
flux from these observations. 
Similar intermediate BL~Lac--QSO behaviour
is not unique to \object{B2\,1308+326}, and has also been observed in
\object{PKS\,0521-365} for example \cite{1999ApJ...526..643S}.

The existence of sources intermediate between BL~Lac and QSOs is
expected in the blazar unification paradigm presented by Fossati~et~al.\
\cite*{1998MNRAS.299..433F}, where synchrotron peak frequency is correlated
with source luminosity.  X-ray, optical and radio luminosities for
\object{B2\,1308+326} are compared to averages for the different blazar
samples examined by Fossati~et~al.\ \cite*{1998MNRAS.299..433F} in
Table~\ref{blazar_lumin}.  It is interesting to note that the luminosities
for \object{B2\,1308+326} span the range between the RBL-FSRQ averages.
\begin{table}
\caption{Average radio, optical and x-ray luminosities
	 for samples of XBLs (Slew), RBLs (1\,Jy) and FSRQs.  The Slew,
	 1\,Jy and FSRQ sample data (Cols.~1, 2 and 3 respectively) are from
	 Fossati~et~al.\ \protect\cite*{1998MNRAS.299..433F}.  Luminosities
	 calculated from the contemporaneous data for B2\,1308+326 are shown
	 for comparison.  The 5\,GHz luminosity of B2\,1308+326 is estimated
	 from an extrapolation of the best-fit parabola to the
	 contemporaneous data.  The minimum recorded flux at 5\,GHz and the
	 maximum recorded flux in the V-band for \object{B2\,1308+326} are
	 shown in parentheses.  Luminosities are given in units of erg
	 s$^{-1}$.}
\label{blazar_lumin}
\begin{tabular}{@{}l c c c c@{}}
\hline\hline
Band	& XBLs (Slew)	& RBLs (1\,Jy)	& FSRQ	& \object{B2\,1308+326}\\
\hline
5\,GHz	& 41.71	& 43.69	& 44.81	& 45.4 (44.0)\\
V-band	& 44.91	& 45.68	& 46.58	& 45.1 (46.7)\\
1\,keV	& 44.94	& 44.72	& 45.98	& 44.9\\
\hline
\end{tabular}
\end{table}

Sources intermediate between BL~Lac and QSOs
with luminous emission lines that are SSC-dominated
and have high synchrotron peak frequencies \cite{1998AJ....115.1253P,2000ApJ...533..650S} still
pose serious problems to the unification scheme proposed by Fossati~et~al.\
\cite*{1998MNRAS.299..433F}.  It has recently been pointed out
\cite{Georganopoulos:2000} that the observed differences between FSRQs with
high synchrotron peak frequencies (``blue quasars'') and HBLs can be
explained by taking into account the orientation of the jet as well as the
intrinsic source luminosity.  In a similar way earlier unification schemes
have attempted to unify intrinsically lower luminosity FR\,I radio galaxies
with BL~Lacs and the higher luminosity FR\,II radio galaxies with quasars
\cite{1992MNRAS.257..404P,1993ApJ...407...65G}.  It now seems
likely that quasars and BL~Lacs form two distinct populations of higher and
lower luminosity AGN respectively, where the orientation of the jet
influences the luminosity and peak synchrotron frequency.  The unusual
intermediate properties of \object{B2\,1308+326} make it an ideal candidate
for studying the overlap between the BL~Lac and QSO categories.  It may
therefore represent the prototype of this kind of composite source.

\subsection{Gravitational microlensing}
The gravitational microlensing scenario
\cite{1993ApJ...410...39G,1987Natur.330..348M}, provides an
alternate explanation for the BL~Lac properties of
\object{B2\,1308+326}. Its high bolometric luminosity is
comparable to that of other gravitationally lensed candidates
\object{AO\,0235+164} \cite{1996A&A...310....1R} and
\object{PKS\,0537-441}
\cite{1999A&AS..135..477R,1996ApJS..104..251P} and may be due to
amplification by a foreground galaxy. \ion{Mg}{ii} absorption
lines with very small W$_{\lambda}$ ($\sim0.4$\,\AA) were reported
\cite{Miller:78} at $z=0.879$ in \object{B2\,1308+326}.  These
lines have not been confirmed by later observations.
Recent imaging with HST \cite{1999ApJ...512...88U} did not
confirm the detection of spatially extended emission
\cite{1993A&AS...98..393S} and no evidence was found
for a foreground lensing galaxy.  Such a galaxy is therefore
limited to an I-band magnitude greater than 20.1, which, if the
galaxy were at a redshift $z=0.879$ implies M$_{\mathrm I}\gtrsim-23$, thus not constraining its luminosity
significantly.  The absorbing column towards
\object{B2\,1308+326} is greater than that expected from our
Galaxy (Fig.~\ref{1308contours}).  The excess absorption could
be due to an absorber at $z<0.997$ or equally, to absorption in
the host galaxy of \object{B2\,1308+326}. The detection of excess
absorption therefore does not exclude the possibility of a
foreground lensing galaxy.  A more detailed spectroscopic
observation with XMM could determine the redshift of the
foreground absorber.  Sensitive, high-resolution optical
spectroscopy could confirm the detection of \ion{Mg}{ii}
absorption lines \cite{Miller:78} and thus also determine the
redshift of such an absorber.

\subsection{Constraints on beaming in \object{B2\,1308+326}}
\begin{table*}
\caption{Comparison of relativistic jet parameters calculated using
     contemporaneous data compared to those of Ghisellini~et~al.
     \protect\cite*{1993ApJ...407...65G} and Ter\"{a}sranta \& Valtaoja
     \protect\cite*{1994A&A...283...51T}.
     $F_m$ is the flux at the synchrotron self-absorption frequency
     $\nu_m$, $F_x$ is the x-ray or optical flux at the energy $\nu_x$,
     \,$\nu_b$ is the synchrotron high energy cut-off frequency, $\alpha$
     is the spectral index of the thin synchrotron emission, $\beta_a$
     is the highest superluminal speed, $\theta_d$ is the VLBI FWHM core
     size at $\nu_m$, $z$ is the source redshift, $\delta$ is a lower
     limit on the Doppler boost factor, T$_{\rm B}$ is the observed
     brightness temperature, $\phi$ is an upper limit on the angle
     between the jet velocity and the line of sight and $\Gamma$ is a
     lower limit on the Lorentz factor of the jet. $\delta$, T$_{\rm
     B}$, $\phi$ and $\Gamma$ were calculated using the data listed in
     that row.  The third and fourth rows use data from observations
     made contemporaneously with the ASCA observations using the two
     different VLBI core sizes reported by Gabuzda~et~al.
     \protect\cite*{1993ApJ...410...39G}.}
\label{relfactors}
\setlength{\extrarowheight}{3pt}
\addtolength{\tabcolsep}{-1pt}
\begin{tabular}{@{}l l l l l l l l l l l l c@{}}
\hline
\hline
$F_m$   & $\nu_m$   & $F_x$         & $\nu_x$       & $\nu_b$   & $\alpha$  & $\beta_a^{Ga}$        & $\theta_d$            & $z$   & $\delta$
    & T$_{\rm B}$       & $\phi$    & $\Gamma$  \\
(Jy)    & (GHz)     & (Jy)          & (keV)         & (GHz)     &       &               & (mas)             &   &                                                                                                                   & ($10^{12}$\,K)    & ($\degr$) &       \\
\hline
1.97$^{Gh}$& 5.0$^{Gh}$    & $0.3\times10^{-6}$\,$^{Gh}$&1$^{Gh}$      & $1\times10^5$\,$^{Gh}$& 0.75$^{Gh}$& 15.75    & 0.5$^{Gh}$            & 0.996$^{Gh}$& 6.8$^{Gh}$  & 1.11$^{Gh}$       & 6.1       & 21.7      \\
--  & --        & --            & --            & --        & --        & 15.75     & --                & 0.992$^T$& 1.9$^T$    & 6.40$^T$      & 7.2       & 68.0      \\
1.22$^{Ga}$& 5.0$^{Ga}$ & $0.18\times10^{-3}$   & $2.28\times10^{-3}$   & $1\times10^3$ & 0.80      & 15.75     & 0.05$^{Ga}$           & 0.997 & 122       & 69.0          & 0.12      & 62.0      \\
0.76$^{Ga}$& 5.0$^{Ga}$& $0.18\times10^{-3}$    & $2.28\times10^{-3}$   & $1\times10^3$ & 0.80      & 15.75     & 0.18$^{Ga}$           & 0.997 & 9.4       & 3.35          & 5.37      & 18.0      \\
\hline

\end{tabular}
\vspace{2mm}

\noindent\makebox[1.8em][r]{$^{Gh}$}\footnotesize{Data from Ghisellini~et~al.\ \cite*{1993ApJ...407...65G}}
\noindent\makebox[1.8em][r]{$^{Ga}$}\footnotesize{Data from Gabuzda~et~al.\ \cite*{1993ApJ...410...39G}}
\noindent\makebox[1.8em][r]{$^T$}\footnotesize{Data from Ter\"{a}sranta \& Valtaoja \cite*{1994A&A...283...51T}}
\end{table*}

Contemporaneous multiwavelength observations allow the determination of
a lower limit on the doppler boost factor ($\delta$),
the corresponding observed brightness temperature T$_{\rm B}$, the
maximum angle ($\phi$) between the line of sight and the velocity
of the relativistic jet as well as the minimum Lorentz factor of
the jet, $\Gamma$.  Ghisellini~et~al.\ \cite*{1993ApJ...407...65G}
have derived these values for \object{B2\,1308+326} with data from
different epochs, using the equations:
\begin{equation}
\delta = (1+z)\,f(\alpha)\,F_m\left(\frac{\ln{(\nu_b/\nu_m)}}{F_x\,\theta_d^{6+4\alpha}\,\nu_x^\alpha\,\nu_m^{5+3\alpha}}\right)^\frac{1}{4+2\alpha}
\label{doppler}
\end{equation}
\begin{equation}
{\rm T}_{\rm B}=1.77\times10^{12}(1+z)\frac{F_m}{\theta_d^2\nu_m^2}\,{\rm K}
\label{temperature}
\end{equation}
\begin{equation}
\phi=\arctan{\frac{2\beta_a}{\beta_a^2+\delta^2-1}}
\label{phi}
\end{equation}
and
\begin{equation}
\Gamma=\frac{\beta_a^2+\delta^2+1}{2\delta},
\label{Lorentz}
\end{equation}
where $F_m$ and $F_x$ are the fluxes in Jy at the synchrotron
self-absorption frequency, $\nu_m$ (GHz), and at the x-ray or
optical energy, $\nu_x$ (keV) respectively;  $\nu_b$ is the
synchrotron high energy cut-off frequency;  $\alpha$ is the
spectral index of the thin synchrotron emission defined from
$F_\nu\propto\nu^{-\alpha}$;  $\beta_a$ is the fastest reported
apparent superluminal speed;  $\theta_d$ is the VLBI FWHM core
size \cite{1977ApJ...216..244M} in milliarcseconds (mas) at
$\nu_m$; $z$ is the redshift of the source.  A comparison of these
values obtained at different epochs for \object{b2\,1308+326} is made in
Table~\ref{relfactors}.

It has been noted that the largest source of error in
Eq.~(\ref{doppler}) is the determination of the VLBI core
size, $\theta_d$ \cite{1987MNRAS.224..257M}.  This value is
crucial in determining the limit on $\delta$.

Ghisellini~et~al.\ \cite*{1993ApJ...407...65G} calculate
$\delta=6.8$ using $\theta_d=0.5$\,mas at 5\,GHz.  High
resolution VLBI images of \object{B2\,1308+326} were presented by
Gabuzda~et~al.\ \cite*{1993ApJ...410...39G} obtained in 1987 and
1989.  Using the 5\,GHz fluxes and corresponding FWHM core sizes
of Gabuzda~et~al.\ \cite*{1993ApJ...410...39G}, 
$\delta$ values of $122$ and $9.4$ were determined with the data
presented here. Clearly the core size is variable and further
constraints on $\delta$ would require VLBI imaging
simultaneous with observations at other wavelengths.

Ter\"{a}sranta \& Valtaoja \cite*{1994A&A...283...51T} use the
radio variability timescales of \object{B2\,1308+326} to deduce a lower
limit for the doppler boost factor of $\delta=1.9$ and the
corresponding maximum observed brightness temperature
T$_{\rm B,max}=6.4\times10^{12}$\,K.
The 22\,GHz light curve used for this calculation is plotted in
Fig.~\ref{radio}.  This method provides a useful
independent measure of $\delta$.  A comparison of these values
with those derived using Eqs.~(\ref{doppler})--(\ref{Lorentz})
are given in Table~\ref{relfactors}.  Two sets of values have been
derived with the data obtained contemporaneously with ASCA using the
different core sizes reported by Gabuzda~et~al.
\cite*{1993ApJ...410...39G}.  The contemporaneous data indicates that
the synchrotron peak frequency is roughly $10^{12}$\,Hz (Fig.~\ref{broad})
and this value was used for $\nu_b$. A similar fit to the highest archival
fluxes in each waveband and the ROSAT data below the
spectral break yields a peak synchrotron frequency of $\sim10^{14}$\,Hz.
The V-band flux and frequency were used instead of the 1\,keV flux in
these calculations, as the 1\,keV flux at the time of these observations was
almost certainly dominated by inverse Compton emission.

Three of the limits on $\delta$ agree within a factor of five
(Table~\ref{relfactors}).  The extreme value of $\delta=122$, obtained
using the core size $\theta_d=0.05$\,mas, is unlikely given the low
flux state of the source during these observations.  While the upper
limits on $\phi$ are consistent with expected jet angles for RBLs,
they are also consistent with those of highly-polarised quasars
(HPQs; Ghisellini~et~al.\ 1993\nocite{1993ApJ...407...65G}, Urry
\& Padovani 1995\nocite{1995PASP..107..803U}). Quasars generally
appear to have larger values of $\delta$ than BL~Lacs
\cite{1993ApJ...410...39G,1994A&A...283...51T}.  A mean $\delta$
of 8.3 for HPQs and 1.3 for BL~Lacs \cite{1993ApJ...410...39G}
indicates that \object{B2\,1308+326} with $\delta=9.4$, is
rather extreme for a typical BL~Lac.

\section{Conclusions}
The blazar \object{B2\,1308+326} was observed contemporaneously at x-ray,
optical and radio wavelengths in June 1996.  The source was in a low
synchrotron flux state at the time of observation and no variability was
detected.  The ROSAT data reveal an x-ray spectrum that is best fit by a
broken power law with absorber model, with the break at $1.1$\,keV in the
rest-frame of the blazar.  The break is probably due to the emerging
dominance of the IC over the synchrotron component.  The frequency of the
peak synchrotron emission component appears to have decreased with
decreasing flux, thus indicating that SSC may be the dominant emission
mechanism in this source.  The x-ray IC component is unaffected by the
large change in the synchrotron emission. \ion{Mg}{ii} emission was observed
with rest W$_{\lambda}\sim$\,15\AA, significantly different from the
W$_{\lambda}$ values previously reported.  The variable W$_{\lambda}$
emission in \object{B2\,1308+326} is probably mostly due to the highly
variable continuum.  Although values of W$_{\lambda}$ are quite small, the
line luminosity is only slightly lower than in a sample of QHBs.
A lower limit on the Doppler boost factor obtained from the
contemporaneous data is consistent with expectations for HPQs but
is higher than expected for BL~Lacs. X-ray absorption at a
level in excess of the Galactic value was detected and indicates
the possible presence of a foreground absorber.

\object{B2\,1308+326} is a typical RBL in terms of peak
synchrotron power and optical variability, but with a very large bolometric
luminosity, variable line emission and a high Doppler boost factor also has
quasar-like attributes.  It may be an intermediate or transitional AGN or a
gravitationally microlensed quasar, given its absorption above the Galactic
level. Future x-ray observations could determine the nature of the absorber.
Further simultaneous multiwavelength observations of this possibly
prototypical intermediate blazar could determine the cause of both its
BL~Lac and its quasar-like properties, providing insight into the relation
between BL~Lacs and FSRQs.

\paragraph{Acknowledgement.}  We would like to thank
A.~Celotti, for helpful comments that resulted in significant
improvements to the paper.

\bibliography{mnemonic,/users/dwatson/work/latex/refs/refs}
\end{document}